\documentstyle[prd,aps,preprint,floats,axodraw]{revtex}

\tightenlines

\input epsf.tex

\renewcommand{\Re}{{\cal R}e}
\renewcommand{\Im}{{\cal I}m}
\newcommand{\Arg}{\text{Arg}}

\newcommand{\phicp}{\phi_{\text{CP}}}
\newcommand{\amu}{a_{\mu}}
\newcommand{\dmu}{d_{\mu}}
\newcommand{\de}{d_e}
\newcommand{\amusm}{a_{\mu}^{\text{SM}}}
\newcommand{\amunp}{a_{\mu}^{\text{NP}}}
\newcommand{\dmusm}{d_{\mu}^{\text{SM}}}
\newcommand{\dmunp}{d_{\mu}^{\text{NP}}}
\newcommand{\amuexp}{a_{\mu}^{\text{exp}}}

\newcommand{\Bino}{\tilde{B}}

\newcommand{\msel}{m_{\tilde{e}}}
\newcommand{\msmu}{m_{\tilde{\mu}}}
\newcommand{\tb}{\tan\beta}

\newcommand{\postscript}[2]{\setlength{\epsfxsize}{#2\hsize}
   \centerline{\epsfbox{#1}}}

\newcommand{\ecm}{e~\text{cm}}

\newcommand{\gev}{\text{GeV}}
\newcommand{\tev}{\text{TeV}}

\newcommand{\eg}{{\em e.g.}}

\newcommand{\eqref}[1]{Eq.~(\ref{#1})}

\newcommand{\bold}[1]{{\text{\normalsize\boldmath $#1$}}}

\begin{document}

\draft

\renewcommand{\thefootnote}{\fnsymbol{footnote}}
\setcounter{footnote}{0}

\preprint{
\hfil
\begin{minipage}[t]{3in}
\begin{flushright}
MIT--CTP--3137\\
UCI--TR--2001--17\\
CERN--TH/2001--118\\
WIS/8/01--May--DPP\\ 
hep-ph/0107182
\end{flushright}
\end{minipage}
}

\title{
\vspace*{.5in}
Theoretical Expectations for the Muon's Electric Dipole Moment}
\author{
Jonathan L.~Feng$^{ab}$,
Konstantin T.~Matchev$^{c}$,
and Yael Shadmi$^{d}$
\vskip 0.3in
}

\address{
  ${}^{a}$
  Center for Theoretical Physics,
  Massachusetts Institute of Technology\\
  Cambridge, MA 02139, U.S.A.\\ 
\vskip 0.15in
  ${}^{b}$
  Department of Physics and Astronomy,
  University of California, Irvine\\
  Irvine, CA 92697, U.S.A.\\
\vskip 0.15in
  ${}^{c}$
  Theory Division, CERN\\
  CH--1211, Geneva 23, Switzerland\\
\vskip 0.15in
  ${}^{d}$
  Department of Particle Physics,
  Weizmann Institute of Science\\
  Rehovot 76100, Israel
}


\maketitle

\begin{abstract} 
We examine the muon's electric dipole moment $\dmu$ from a variety of
theoretical perspectives.  We point out that the reported deviation in
the muon's $g-2$ can be due partially or even entirely to a new
physics contribution to the muon's {\em electric} dipole moment.  In
fact, the recent $g-2$ measurement provides the most stringent bound
on $\dmu$ to date.  This ambiguity could be definitively resolved by
the dedicated search for $\dmu$ recently proposed.  We then consider
both model-independent and supersymmetric frameworks. Under the
assumptions of scalar degeneracy, proportionality, and flavor
conservation, the theoretical expectations for $\dmu$ in supersymmetry
fall just below the proposed sensitivity. However, non-degeneracy can
give an order of magnitude enhancement, and lepton flavor violation
can lead to $\dmu \sim 10^{-22}~\ecm$, two orders of magnitude above
the sensitivity of the $\dmu$ experiment.  We present compact
expressions for leptonic dipole moments and lepton flavor violating
amplitudes.  We also derive new limits on the amount of flavor
violation allowed and demonstrate that approximations previously used
to obtain such limits are highly inaccurate in much of parameter
space.
\end{abstract}



\renewcommand{\thefootnote}{\arabic{footnote}}
\setcounter{footnote}{0}

\newpage

\section{Introduction}
\label{sec:intro}

Electric dipole moments (EDMs) of elementary particles are predicted
to be far below foreseeable experimental sensitivity in the standard
model.  In extensions of the standard model, however, much larger EDMs
are possible.  Current EDM bounds are already some of the most
stringent constraints on new physics, and they are highly
complementary to many other low energy constraints, since they require
$CP$ violation, but not flavor violation.

The field of precision muon physics will be transformed in the next
few years~\cite{Hawaiiproc}.  The EDM of the muon is therefore of
special interest. A new experiment~\cite{Semertzidis:1999kv} has been
proposed to measure the muon's EDM at the level of
\begin{equation}
\dmu \sim 10^{-24}~\ecm \ ,
\label{proposedEDM}
\end{equation}
more than five orders of magnitude below the current
bound~\cite{Bailey:1979mn}
\begin{equation}
\dmu = (3.7 \pm 3.4) \times 10^{-19}~\ecm \ .
\label{currentEDM}
\end{equation}
The interest in the muon's EDM is further heightened by the recent
measurement of the muon's anomalous magnetic dipole moment (MDM) $\amu
= (g_{\mu}-2)/2$, where $g_{\mu}$ is the muon's gyromagnetic ratio.
The current measurement $\amuexp = 11\ 659\ 202\, (14)\, (6) \times
10^{-10}$~\cite{Brown:2001mg} from the Muon $(g-2)$ Experiment at
Brookhaven differs from the standard model prediction
$\amusm$~\cite{Davier:1998si,Marciano:2001qq} by $2.6 \sigma$:
\begin{equation}
\Delta a_\mu \equiv \amuexp - \amusm = (43 \pm 16) \times 10^{-10} \ .
\label{currentamu}
\end{equation}
The muon's EDM and $\amu$ arise from similar operators, and this
tentative evidence for a non-standard model contribution to $\amu$
also motivates the search for the muon's EDM.

In this study, we examine the prospects for detecting a non-vanishing
muon EDM from a variety of theoretical perspectives.  We first note
that the reported deviation in the muon's $g-2$ can be due partially
or even entirely to a new physics contribution to the muon's {\em
electric} dipole moment.  In fact, at present the result from the Muon
$(g-2)$ Experiment provides the most stringent bound on $\dmu$.  We
derive this bound and comment on the conclusions that may be drawn
about $\dmu$ from the $\amu$ measurement alone.

We then move to more concrete frameworks, where additional
correlations constrain our expectations.  In particular, we consider
supersymmetry and examine quantitatively the implications of the
electron EDM and lepton flavor violating
processes~\cite{recentwork,Ibrahim:2001jz}.  Our aim is to impose as
few theoretical prejudices as possible and draw correspondingly
general conclusions.  For studies of the muon EDM in specific
supersymmetric models, see, \eg,
Refs.~\cite{Babu:2000dq,Blazek:2001zm}.

Finally, although we use exact expressions for all flavor-conserving
amplitudes in this study, we also provide compact expressions in the
mass insertion approximation for branching ratios of radiative lepton
decays and for leptonic EDMs and MDMs both with and without lepton
flavor violation.  These include all leading supersymmetric effects
and are well-suited to numerical calculations.

\section{Model-independent bounds from the Muon 
$\bold{(\lowercase{g}-2)}$ Experiment}
\label{sec:experimental}

Modern measurements of the muon's MDM exploit the equivalence of
cyclotron and spin precession frequencies for $g=2$ fermions
circulating in a perpendicular and uniform magnetic field.
Measurements of the anomalous spin precession frequency are therefore
interpreted as measurements of $\amu$.

The spin precession frequency also receives contributions from the
muon's EDM, however. For a muon traveling with velocity $\bold{\beta}$
perpendicular to both a magnetic field $\bold{B}$ and an electric
field $\bold{E}$, the anomalous spin precession vector is
\begin{equation}
\bold{\omega}_a = -a_{\mu} \frac{e}{m_{\mu}} \bold{B}
- d_{\mu} \frac{2c}{\hbar} \bold{\beta} \times \bold{B} 
- \frac{e}{m_{\mu}c} \left(\frac{1}{\gamma^2-1} - a_{\mu}\right) 
\bold{\beta} \times \bold{E} 
- d_{\mu} \frac{2}{\hbar} \bold{E} \ .
\label{omega}
\end{equation}
In recent experiments, the third term of \eqref{omega} is removed by
running at the `magic' $\gamma \approx 29.3$, and the last term is
negligible.  For highly relativistic muons with $|\bold{\beta}|
\approx 1$, then, the anomalous precession frequency is
\begin{equation}
|\bold{\omega}_a| 
\approx |\bold{B}| \left[ \left( \frac{e}{m_{\mu}} \right)^2
\left({\amusm}^2 + 2 \amusm \amunp \right) + 
\left(\frac{2c}{\hbar}\right)^2 {\dmunp}^2 \right]^{1/2} \ ,
\label{both}
\end{equation}
where NP denotes new physics contributions, and we have assumed
$\amunp \ll \amusm$ and $\dmunp \gg \dmusm$.

\begin{figure}[tbp]
\postscript{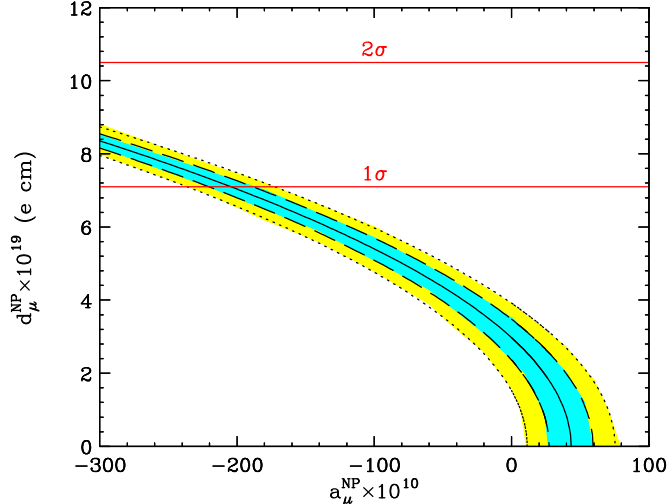}{0.53}
\caption{Regions in the $(\amunp,\dmunp)$ plane that are consistent
with the observed $|\bold{\omega}_a|$ at the 1$\sigma$ and 2$\sigma$
levels.  The current 1$\sigma$ and 2$\sigma$ bounds on
$\dmunp$~\protect\cite{Bailey:1979mn} are also shown.}
\label{fig:amu_dmu}
\end{figure}

The observed deviation from the standard model prediction for
$|\bold{\omega}_a|$ has been assumed to arise entirely from a MDM and
has been attributed to a new physics contribution of size $\Delta
a_\mu$.  However, from \eqref{both}, we see that, more generally, it
may be due to some combination of magnetic and electric dipole moments
from new physics.  More quantitatively, the effect can also be due to
an EDM contribution
\begin{equation}
\left| \dmunp \right| \approx \frac{\hbar e}{m_{\mu} c} \,
\sqrt{\ \frac{1}{2}\, \amusm\, 
\left(\Delta a_\mu - \amunp  \right)}
\approx 3.0 \times 10^{-19}~\ecm\ 
\sqrt{1 - \frac{\amunp}{43 \times 10^{-10}}} \ ,
\label{mdmisedm}
\end{equation}
where $\amunp$ has been normalized to the current central value given
in \eqref{currentamu}. In Fig.~\ref{fig:amu_dmu} we show the regions
in the $(\amunp,\dmunp)$ plane that are consistent with the observed
deviation in $|\bold{\omega}_a|$.

In fact, the observed anomaly may, in principle, be due entirely to
the muon's EDM! This is evident from Eqs.~(\ref{currentEDM}) and
(\ref{mdmisedm}), or from Fig.~\ref{fig:amu_dmu}, where the current
1$\sigma$ and 2$\sigma$ upper bounds on $\dmunp$~\cite{Bailey:1979mn}
are also shown.  Alternatively, in the absence of fine-tuned
cancellations between $\amunp$ and $\dmunp$, {\em the results of the
Muon $(g-2)$ Experiment also provide the most stringent bound on
$\dmu$ to date}, with $1\sigma$ and $2\sigma$ upper limits
\begin{equation}
\Delta a_{\mu} < 59 \ (75) \times 10^{-10} \Longrightarrow
\left| \dmunp \right| < 3.5 \ (3.9) \times 10^{-19}~\ecm \ .
\label{newdmubound}
\end{equation}

This discussion is completely model-independent. In specific models,
however, it may be difficult to achieve values of $\dmu$ large enough
to saturate the bound of \eqref{newdmubound}.  For example, in
supersymmetry, assuming flavor conservation and taking extreme values
of superparticle masses ($\sim 100~\gev$) and $\tb$ ($\tb \sim 60$) to
maximize the effect, the largest possible value of $\amu$ is
$a_{\mu}^{\text{max}} \sim 10^{-7}$~\cite{Feng:2001tr}.  Very roughly,
one therefore expects a maximal muon EDM of order $(e \hbar / 2
m_{\mu} c) a_{\mu}^{\text{max}} \sim 10^{-20}~\ecm$ in supersymmetry.

Of course, the effects of $\dmu$ and $\amu$ are physically
distinguishable: while $\amu$ causes precession around the magnetic
field's axis, $\dmu$ leads to oscillation of the muon's spin above and
below the plane of motion.  This oscillation is detectable in the
distribution of positrons from muon decay, and further analysis of the
recent $\amu$ data should tighten the current bounds on $\dmu$.  Such
analysis is currently in progress~\cite{lee}.  The proposed dedicated
$\dmu$ search will provide a definitive answer, however, by either
measuring a non-zero $\dmu$ or constraining the contribution of $\dmu$
to $|\bold{\omega}_a|$ to be insignificant.

\section{Theoretical Expectations from the muon's MDM}
\label{sec:theoretical}

\begin{figure}[tbp]
\postscript{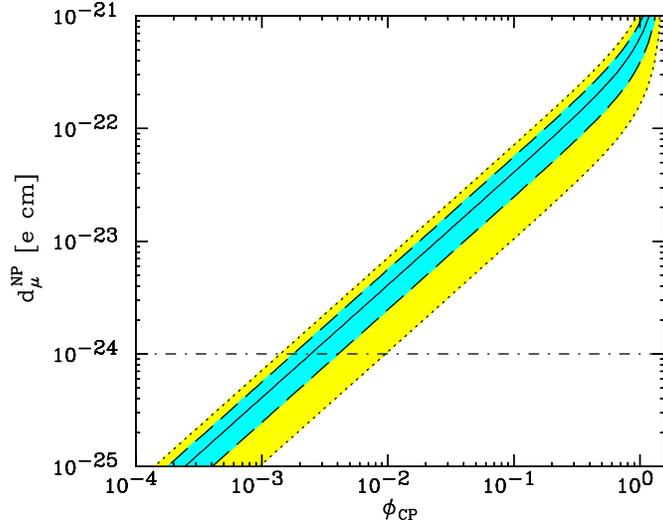}{0.53}
\caption{Regions of the $(\phicp, \dmunp)$ plane allowed by the
measured central value of $|\bold{\omega}_a|$ (solid) and its
1$\sigma$ and 2$\sigma$ preferred values (shaded).  The horizontal
line marks the proposed experimental sensitivity to $\dmunp$. }
\label{fig:dmu_phi}
\end{figure}

The muon's EDM and anomalous MDM are defined through\footnote{Here and
below we set $\hbar = c = 1$.}
\begin{eqnarray}
\label{EDMoperator}
{\cal L}_{\text{EDM}} &=& 
-\frac{i}{2} \dmunp \, \bar{\mu} \sigma^{mn} \gamma_5 \mu \, F_{mn} \\ 
{\cal L}_{\text{MDM}} &=& 
\amunp \frac{e}{4m_\mu} \, \bar{\mu} \sigma^{mn} \mu \, F_{mn} \ ,
\end{eqnarray}
where $\sigma^{mn} = \frac{i}{2} \left[ \gamma^m, \gamma^n \right]$
and $F$ is the electromagnetic field strength.  These operators are
closely related.  In the absence of all other considerations, one
might expect their coefficients to be of the same order.
Parameterizing them as $\dmunp/2 = \Im A$ and $\amunp e/(4 m_{\mu}) =
\Re A$ with $A \equiv |A|e^{i\phicp}$, one finds
\begin{equation}
\dmunp = 4.0 \times 10^{-22}~\ecm\ \frac{\amunp}{43 \times 10^{-10}} 
\ \tan\phicp \ .
\label{phicp}
\end{equation}
The measured discrepancy in $|\bold{\omega}_a|$ then constrains
$\phicp$ and $\dmunp$.  The preferred regions of the $(\phicp,
\dmunp)$ plane are shown in Fig.~\ref{fig:dmu_phi}.  For `natural'
values of $\phicp \sim 1$, $\dmunp$ is of order $10^{-22}~\ecm$.  With
the proposed $\dmunp$ sensitivity of \eqref{proposedEDM}, all of the
2$\sigma$ allowed region with $\phicp > 10^{-2}$ yields an observable
signal.

At the same time, while this model-independent analysis indicates that
natural values of $\phicp$ prefer $\dmunp$ well within reach of the
proposed muon EDM experiment, very large values of $\dmunp$ also
require highly fine-tuned $\phicp$.  For example, the contributions of
$\dmunp$ and $\amunp$ to the observed discrepancy in $\amu$ are
roughly equal only if $|\pi/2 - \phicp| \sim 10^{-3}$.  This is a
consequence of the fact that EDMs are $CP$-odd and $\dmusm \approx 0$,
and so $\dmunp$ appears only quadratically in $|\bold{\omega}_a|$.
Without a strong motivation for $\phicp \approx \pi/2$, it is
therefore natural to expect the EDM contribution to
$|\bold{\omega}_a|$ to be negligible, and we assume in the following
that the $|\bold{\omega}_a|$ measurement is indeed a measurement of
$\amu$.

\section{The electron EDM and naive scaling}
\label{sec:scaling}

The EDM operator of \eqref{EDMoperator} couples left- and right-handed
muons, and so requires a mass insertion to flip the chirality.  The
natural choice for this mass is the lepton mass.  On dimensional
grounds, one therefore expects
\begin{equation}
\label{massscaling} 
\dmunp \propto \frac{m_{\mu}}{\tilde{m}^2}\ , 
\end{equation} 
where $\tilde{m}$ is the mass scale of the new physics. If the new
physics is flavor blind, $d_f \propto m_f$ for all fermions $f$, which
we refer to as `naive scaling.'  In particular,
\begin{equation}\label{naive}
d_\mu \approx {m_\mu\over m_e}\, d_e \ .
\end{equation}

The current bound on the electron EDM is $d_e = 1.8\, (1.2)\, (1.0)
\times 10^{-27}~\ecm$~\cite{Commins:1994gv}.  Combining the
statistical and systematic errors in quadrature, this bound and
\eqref{naive} imply
\begin{equation}
d_\mu \alt 9.1\times 10^{-25}~\ecm \ ,
\label{muedmlimit}
\end{equation}
at the 90\% CL, which is barely below the sensitivity of
\eqref{proposedEDM}.  Naive scaling must be violated if a
non-vanishing $\dmu$ is to be observable at the proposed experiment.
On the other hand, the proximity of the limit of \eqref{muedmlimit} to
the projected experimental sensitivity of \eqref{proposedEDM} implies
that even relatively small departures from naive scaling may yield an
observable signal.

\section{Violations of naive scaling in supersymmetry}
\label{sec:violations}

Is naive scaling violation well-motivated, and can the violation be
large enough to produce an observable EDM for the muon? To investigate
these questions quantitatively, we consider supersymmetry. (For
violations of naive scaling in other models, see, for example,
Ref.~\cite{Babu:2000cz}.) Many additional mass parameters are
introduced in supersymmetric extensions of the standard model. These
are in general complex and so are new sources of $CP$ violation.
These parameters may be correlated by a fundamental theory of
supersymmetry breaking that includes a specific mechanism for
mediating the breaking.  In fact, all viable mechanisms of mediating
supersymmetry breaking are designed to suppress flavor violation, and
so $CP$-violating observables that also involve flavor violation, such
as $\epsilon_K$, are also suppressed.  However, EDMs do not require
flavor violation, and constraints on quark and electron EDMs are some
of the main challenges for supersymmetric models.  For a recent
discussion of the supersymmetric $CP$ problem in various supersymmetry
breaking schemes, see Ref.~\cite{Dine:2001ne}.

In full generality, the relevant dimensionful supersymmetry parameters
for leptonic EDMs are the slepton mass matrices $\bold{m}^2$,
trilinear scalar couplings $\bold{A}$, gaugino masses $M_1$ and $M_2$,
the Higgsino mass $\mu$, and the dimension two Higgs scalar coupling
$B$. Schematically, these enter the Lagrangian through the terms
\begin{eqnarray}\label{soft}
{\cal L} &\supset& 
\bold{m}_{LL}^2{}_{ij} \tilde{L}^*_i \tilde{L}_j ,\,
\bold{m}_{RR}^2{}_{ij} \tilde{E}^*_i \tilde{E}_j ,\,
\bold{A}_{ij} H_d \tilde{L}_i \tilde{E}_j ,\, 
M_1 \tilde{B} \tilde{B} ,\, M_2 \tilde{W} \tilde{W} ,\, 
\mu \tilde{H}_u \tilde{H}_d ,\, B H_u H_d \ ,
\end{eqnarray}
where $i,j$ are generational indices, $L$ and $E$ denote SU(2) doublet
and singlet leptons, respectively, $H_u$ and $H_d$ are the up- and
down-type Higgs multiplets, and $\tilde{B}$ and $\tilde{W}$ are
gauginos, the U(1) Bino and SU(2) Winos.  The parameter $\tb = \langle
H_u^0 \rangle / \langle H_d^0 \rangle$, the ratio of Higgs boson
vacuum expectation values, will also enter below.

The U(1)$_R$ and U(1)$_{PQ}$ symmetries allow us to remove two phases
--- we choose $M_2$ and $B$ real.  Throughout this study, we assume
that the gaugino masses have a common phase, as is true in many
well-motivated theories where the gaugino masses are either unified at
some scale or otherwise have a common origin. We also begin by
assuming supersymmetric flavor conservation, that is, that the
sfermion masses and trilinear couplings are diagonal in the fermion
mass basis. (We will consider flavor violation in
Sec.~\ref{sec:flavor}.)  With these assumptions, only the $\mu$ and
$A_{\ell}$ parameters are complex.  We define $\phi_{\mu} \equiv \Arg
(\mu)$ and $\phi_{A_{\ell}} \equiv \Arg (A_{\ell})$.

Leptonic electromagnetic dipole moments arise at one-loop from
chargino-sneutrino and neutralino-charged slepton diagrams. In all
figures and results presented here, we evaluate flavor-conserving
amplitudes exactly.  However, for purposes of exposition, it is
convenient to consider the fermion mass basis and to adopt the mass
insertion approximation for sleptons, neutralinos, and charginos.  (We
have checked that the mass insertion approximation is accurate to
about 5\% in almost all of parameter space, justifying the intuition
derived from this simplification.) In the mass insertion
approximation, for large and moderate $\tb$ and neglecting subdominant
terms, there are five contributions with the following Feynman
diagrams and amplitudes~\cite{Moroi:1996yh}:
\begin{eqnarray}
\parbox{1.5in}{
\begin{picture}(100,60)(0,25)
\Line( 1.0,40.0)(99.0,40.0)
\Text(95.0,33.0)[]{$\ell_R$}
\Text( 5.0,33.0)[]{$\ell_L$}
\Text(50.0,33.0)[]{$\tilde B^0$}
\DashCArc(50.0,40.0)(30.0,0.0,180.0){3.0}
\Text(20.0,75.0)[]{$\tilde \ell_L$}
\Text(80.0,75.0)[]{$\tilde \ell_R$}
\Vertex(50.0,70.0){3}
\Text(50.0,82.0)[]{$\delta^{LR}_{\ell\ell}$}
\end{picture}} \ 
&&
{\cal A}_{\ell}^a = - g'^2 M_1 \left[ A_{\ell} 
\langle H_d^0 \rangle - m_{\ell}\, \mu \tb \right] 
K_N(M_1^2, m_{\tilde{\ell}_L}^2, m_{\tilde{\ell}_R}^2) \nonumber \\
\parbox{1.5in}{
\begin{picture}(100,60)(0,25)
\Line( 1.0,40.0)(99.0,40.0)
\Text(95.0,33.0)[]{$\ell_R$}
\Text( 5.0,33.0)[]{$\ell_L$}
\Text(35.0,33.0)[]{$\tilde H^0$}
\Text(65.0,33.0)[]{$\tilde B^0$}
\DashCArc(50.0,40.0)(30.0,0.0,180.0){3.0}
\Text(50.0,80.0)[]{$\tilde \ell_R$}
\Vertex(50.0,40.0){3}
\end{picture}} \ 
&&
{\cal A}_{\ell}^b = -g'^2 M_1 m_\ell\, \mu \tb\,
K_N(m_{\tilde{\ell}_R}^2, |\mu|^2, M_1^2) \nonumber \\
\parbox{1.5in}{
\begin{picture}(100,60)(0,25)
\Line( 1.0,40.0)(99.0,40.0)
\Text(95.0,33.0)[]{$\ell_R$}
\Text( 5.0,33.0)[]{$\ell_L$}
\Text(65.0,33.0)[]{$\tilde H^0$}
\Text(35.0,33.0)[]{$\tilde B^0$}
\DashCArc(50.0,40.0)(30.0,0.0,180.0){3.0}
\Text(50.0,80.0)[]{$\tilde \ell_L$}
\Vertex(50.0,40.0){3}
\end{picture}} \ 
&&
{\cal A}_{\ell}^c = {1\over2} g'^2 M_1 m_\ell\, \mu \tb\,
K_N(m_{\tilde{\ell}_L}^2, |\mu|^2, M_1^2) 
\nonumber \\
\parbox{1.5in}{
\begin{picture}(100,60)(0,25)
\Line( 1.0,40.0)(99.0,40.0)
\Text(95.0,33.0)[]{$\ell_R$}
\Text( 5.0,33.0)[]{$\ell_L$}
\Text(65.0,33.0)[]{$\tilde H^0$}
\Text(35.0,33.0)[]{$\tilde W^0$}
\DashCArc(50.0,40.0)(30.0,0.0,180.0){3.0}
\Text(50.0,80.0)[]{$\tilde \ell_L$}
\Vertex(50.0,40.0){3}
\end{picture}} \ 
&&
{\cal A}_{\ell}^d = -{1\over2} g_2^2 M_2 m_\ell\, \mu \tb\,
K_N(m_{\tilde{\ell}_L}^2, |\mu|^2, M_2^2) \nonumber \\
\parbox{1.5in}{
\begin{picture}(100,60)(0,25)
\Line( 1.0,40.0)(99.0,40.0)
\Text(95.0,33.0)[]{$\ell_R$}
\Text( 5.0,33.0)[]{$\ell_L$}
\Text(65.0,33.0)[]{$\tilde H^{\pm}$}
\Text(35.0,33.0)[]{$\tilde W^{\pm}$}
\DashCArc(50.0,40.0)(30.0,0.0,180.0){3.0}
\Text(50.0,80.0)[]{$\tilde \nu_\ell$}
\Vertex(50.0,40.0){3}
\end{picture}} \ 
&&
{\cal A}_{\ell}^e = g_2^2 M_2 m_\ell\, \mu \tb\,
K_C(m_{\tilde{\nu}_{\ell}}^2, |\mu|^2, M_2^2) \ . \label{dell} 
\end{eqnarray}
In these diagrams, an external photon connected to any charged
internal line is implicit, and $\delta^{LR}_{\ell\ell} \equiv
(A_{\ell} \langle H_d^0 \rangle - m_{\ell}\, \mu \tb) /
m_{\tilde{\ell}}^2$. The functions $K_N$ and $K_C$ are mass dimension
$-4$ functions entering the neutralino and chargino diagrams,
respectively, and are given in the Appendix.

Defining
\begin{equation}
{\cal A}_{\ell}^{\text{tot}} \equiv {\cal A}_{\ell}^a + 
{\cal A}_{\ell}^b + {\cal A}_{\ell}^c + 
{\cal A}_{\ell}^d + {\cal A}_{\ell}^e \ ,
\label{atot}
\end{equation}
The EDM and anomalous MDM of a lepton $\ell$ are simply
\begin{equation}
d_{\ell} = \frac{1}{2} e \, \Im {\cal A}_{\ell}^{\text{tot}} \ , 
\quad
a_{\ell} = m_{\ell} \, \Re {\cal A}_{\ell}^{\text{tot}} \ .
\label{edmmdm}
\end{equation}

{}From the amplitudes of \eqref{dell}, naive scaling is seen to
require
\begin{itemize}
\item Degeneracy: Generation-independent $m_{\tilde{\ell}_R}$,
$m_{\tilde{\ell}_L}$, and $m_{\tilde{\nu}_{\ell}}$.
\item Proportionality: The $A$ terms must satisfy 
$\Im(A_{\ell}) \propto m_{\ell}$.
\item Flavor conservation: Vanishing off-diagonal elements of
$\bold{m}_{LL}^2$,$\bold{m}_{RR}^2$, and $\bold{A}$.
\end{itemize}
The last requirement, flavor conservation, has been assumed in all of
our discussion so far.  As we will see, relaxing this assumption also
leads to naive scaling violation.  We now consider violations of each
of these properties in turn.

\subsection{Non-degeneracy}
\label{sec:non-degeneracy}

Scalar degeneracy is the most obvious way to reduce flavor changing
effects to allowable levels.  Therefore many schemes for mediating
supersymmetry breaking try to achieve degeneracy.  However, in many of
these, with the exception of simple gauge mediation models,
there may be non-negligible contributions to scalar masses that are
generation-dependent.  Furthermore, there are classes of models that
do not require scalar degeneracy at all.  For example, scalar
non-degeneracy is typical in alignment models~\cite{Nir:1993mx}, where
flavor-changing effects are suppressed by the alignment of scalar and
fermion mass matrices rather than by scalar degeneracy. Scalar
non-degeneracy is also typical in models with anomalous U(1)
contributions to the sfermion masses.  In fact, in models where the
anomalous U(1) symmetry determines both sfermion and fermion masses,
the sfermion hierarchy is often inverted relative to the fermion mass
hierarchy~\cite{Dudas:1996eq,Dudas:1996fe,Brax:2000ip}, and so smuons
are lighter than selectrons, as required for an observable $\dmu$.  In
summary, there is a wide variety of models in which deviations from
scalar degeneracy exist.

We now consider a simple model-independent parameterization to explore
the impact of non-degenerate selectron and smuon masses.  We set
$m_{\tilde{e}_R} = m_{\tilde{e}_L} = \msel$ and $m_{\tilde{\mu}_R} =
m_{\tilde{\mu}_L} = \msmu$ and assume vanishing $A$ parameters.  For
fixed values of $M_1$, $M_2$, $|\mu|$, and large $\tb$, then, to a
good approximation both $\de$ and $\dmu$ are proportional to $\sin
\phi_{\mu}\, \tb$, and we assume that $\sin \phi_{\mu}\, \tb$
saturates the $\de$ bound.

\begin{figure}[tbp]
\postscript{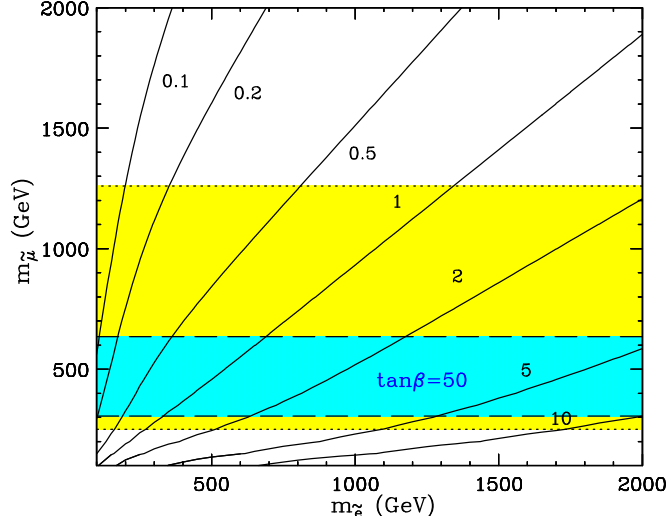}{0.53}
\caption{Contours of $\dmu$ in units of $10^{-24}~\ecm$ for varying
$m_{\tilde{e}_R} = m_{\tilde{e}_L} = \msel$ and $m_{\tilde{\mu}_R} =
m_{\tilde{\mu}_L} = \msmu$ for vanishing $A$ terms, fixed $|\mu| =
500~\gev$ and $M_2 = 300~\gev$, and $M_1 = (g_1^2/g_2^2) M_2$
determined from gaugino mass unification.  The parameter combination
$\sin \phi_{\mu}\, \tb$ is assumed to saturate the bound $\de < 4.4
\times 10^{-27}~\ecm$. The shaded regions are preferred by $\amu$ at
$1\sigma$ and $2\sigma$ for $\tb=50$. }
\label{fig:msel_msmu}
\end{figure}

Contours of $\dmu$ are given in Fig.~\ref{fig:msel_msmu}. The
contributions to $\dmu$ have been evaluated exactly (without the mass
insertion approximation).  Observable values of $\dmu$ are possible
even for small violations of non-degeneracy; for example, for
$\msmu/\msel \alt 0.9$, muon EDMs greater than $10^{-24}~\ecm$ are
possible. The current value of $\amu$ also favors light smuons and
large EDMs.  The smuon mass regions preferred by the current $\amu$
anomaly are given in Fig.~\ref{fig:msel_msmu} for $\tb = 50$.  Within
the $1\sigma$ preferred region, $\dmu$ may be as large as $4\ (10)
\times 10^{-24}~\ecm$ for $\msel < 1\ (2)~\tev$.  Our assumed value of
$\tb$ is conservative; for smaller $\tb$, the preferred smuon masses
are lower and the possible $\dmu$ values larger.

\subsection{Non-proportionality}
\label{sec:prop}

Naive scaling is also broken if the $\Im A_{\ell}$ are not
proportional to Yukawa couplings $y_{\ell}$.  Just as in the case of
non-degeneracy, deviations from proportionality are found in many
models.  Even in models constructed to give proportionality, there are
often corrections to the $A$ terms, so that
\begin{equation}
\bold{A}_{ij}= \bold{y}_{ij} A_0 +\bold{a}_{ij}\ ,
\end{equation}
where the second term is smaller than the first term, but violates
proportionality.  In flavor models, the $A$ terms do
not obey proportionality at all.  Rather they are of the form
\begin{equation}
\bold{A}_{ij}= c_{ij}\, \bold{y}_{ij} A_0\ ,
\end{equation}
where $c_{ij}$ are order one coefficients.  Clearly, violations of
proportionality may affect $d_\mu$ not only by changing the magnitude
of the $A$ terms, but also through the possible appearance of new
phases, either in $\bold{a}_{ij}$ or in $c_{ij}$.  Note that these
possibilities also lead to flavor violation, the subject of the
following section.

We will not study the possibility of non-proportionality in detail.
For large $\tb$, the $A$ term contribution to the EDM is suppressed
relative to the typically dominant chargino contribution by roughly a
factor of $(g_2^2 M_2 / g'^2 M_1) (y_\mu \, \Im\mu\, \tan\beta/\Im
A_\mu)$, where we have used the amplitudes of \eqref{dell}.  However,
there are many possibilities that may yield large effects.  In
Ref.~\cite{Ibrahim:2001jz}, for example, it was noted that $A_e$ may
be such that the chargino and neutralino contributions to $d_e$
cancel, while, since $A_e\neq A_\mu$, there is no cancellation in
$d_\mu$, and observable values are possible.

\subsection{Flavor Violation} 
\label{sec:flavor}

In all of the discussion so far, we have neglected the possibility of
supersymmetric lepton flavor violation.  However, such flavor
violation is present, at least at sub-leading order, in most models of
high-scale supersymmetry breaking~\cite{Dine:2001ne}.  Moreover, large
smuon-stau mixing, of particular importance here, is well-motivated by
the evidence for large $\nu_{\mu}-\nu_{\tau}$ mixing observed in
atmospheric neutrinos~\cite{Fukuda:1998mi}.  Explicit examples of this
connection in models with left-right gauge symmetry are given in
Refs.~\cite{Babu:2000dq,Blazek:2001zm}. The relation between neutrino
and slepton mixing is also a general feature of Abelian flavor
models~\cite{Feng:2000wt}.  In the simplest of these models, highly
mixed states have similar masses, contradicting the most
straightforward interpretation of the neutrino data.  This difficulty
may be circumvented in less minimal models by generating hierarchical
neutrino masses from the neutrino mass matrix and large mixing by
arranging for the gauge and mass eigenstates of the SU(2) lepton
doublets to be related by large rotations.  However, because lepton
doublets contain charged leptons in addition to neutrinos, these
rotations also generate large misalignments between the charged
leptons and sleptons, producing large slepton flavor mixing.

Smuon-stau mixing leads to a potentially significant enhancement in
$\dmu$, because it breaks naive scaling by introducing contributions
with a tau mass insertion so that $\dmu \propto m_{\tau}/\tilde{m}^2$.
However, to evaluate the significance of this enhancement, we must
first determine how large the flavor violation may be.  This effect
may be isolated by assuming that all charged sleptons are roughly
degenerate with characteristic mass $m_{\tilde{\ell}}\/$. In the basis
with lepton mass eigenstates and flavor-diagonal gauge interactions,
slepton flavor violation enters through off-diagonal masses in the
slepton mass matrix. As usual, we parameterize the
chirality-preserving off-diagonal masses by $\delta^{LL}_{23} \equiv
\bold{m}^2_L{}_{23}/m_{\tilde{\ell}}^2$ and $\delta^{RR}_{23} \equiv
\bold{m}^2_E{}_{23}/m_{\tilde{\ell}}^2\/$.  There may also be flavor
violation in the left-right couplings; we parameterize these by
$\delta^{LR}_{23}$ and $\delta^{RL}_{23}$.  We begin by assuming real
$\delta$s; however, very large effects are possible for imaginary
$\delta$s, and we consider this possibility at the end of this
section.

The off-diagonal masses induce $\tau \to \mu \gamma$ transitions.
Eight contributions are parametrically enhanced by $m_{\tau} /
m_{\mu}$ (retaining the possibility that $\delta^{LR,RL}_{23} \propto
m_{\tau}$). Their Feynman diagrams and amplitudes are
\begin{eqnarray}
\parbox{1.5in}{
\begin{picture}(100,60)(0,25)
\Line( 1.0,40.0)(99.0,40.0)
\Text(95.0,33.0)[]{$\tau_L$}
\Text( 5.0,33.0)[]{$\mu_R$}
\Text(50.0,33.0)[]{$\tilde B^0$}
\DashCArc(50.0,40.0)(30.0,0.0,180.0){3.0}
\Text(15.0,55.0)[]{$\tilde \mu_R$}
\Text(88.0,55.0)[]{$\tilde \tau_L$}
\Text(50.0,80.0)[]{$\tilde \tau_R$}
\Vertex(33.0,64.0){3}
\Vertex(67.0,64.0){3}
\Text(22.0,70.0)[]{$\delta^{RR}_{23}$}
\Text(80.0,70.0)[]{$\delta^{LR}_{33}$}
\end{picture}} \ 
&&
\frac{m_{\tau}}{m_{\mu}} \frac{\partial
{\cal A}^a_{\mu}}{\partial \ln m_{\tilde{\mu}_R}^2} \delta^{RR}_{23} 
\qquad 
\parbox{1.5in}{
\begin{picture}(100,60)(0,25)
\Line( 1.0,40.0)(99.0,40.0)
\Text(95.0,33.0)[]{$\tau_R$}
\Text( 5.0,33.0)[]{$\mu_L$}
\Text(50.0,33.0)[]{$\tilde B^0$}
\DashCArc(50.0,40.0)(30.0,0.0,180.0){3.0}
\Text(15.0,55.0)[]{$\tilde \mu_L$}
\Text(88.0,55.0)[]{$\tilde \tau_R$}
\Text(50.0,80.0)[]{$\tilde \tau_L$}
\Vertex(33.0,64.0){3}
\Vertex(67.0,64.0){3}
\Text(22.0,70.0)[]{$\delta^{LL}_{23}$}
\Text(80.0,70.0)[]{$\delta^{RL}_{33}$}
\end{picture}} \ 
\frac{m_{\tau}}{m_{\mu}} \frac{\partial
{\cal A}^a_{\mu}}{\partial \ln m_{\tilde{\mu}_L}^2} \delta^{LL}_{23} 
\nonumber \\
\parbox{1.5in}{
\begin{picture}(100,60)(0,25)
\Line( 1.0,40.0)(99.0,40.0)
\Text(95.0,33.0)[]{$\tau_L$}
\Text( 5.0,33.0)[]{$\mu_R$}
\Text(35.0,33.0)[]{$\tilde B^0$}
\Text(65.0,33.0)[]{$\tilde H^0$}
\DashCArc(50.0,40.0)(30.0,0.0,180.0){3.0}
\Text(20.0,70.0)[]{$\tilde \mu_R$}
\Text(83.0,70.0)[]{$\tilde \tau_R$}
\Vertex(50.0,40.0){3}
\Vertex(50.0,70.0){3}
\Text(50.0,82.0)[]{$\delta^{RR}_{23}$}
\end{picture}} \ 
&&
\frac{m_{\tau}}{m_{\mu}} \frac{\partial
{\cal A}^b_{\mu}}{\partial \ln m_{\tilde{\mu}_R}^2} \delta^{RR}_{23} 
\qquad 
\parbox{1.5in}{
\begin{picture}(100,60)(0,25)
\Line( 1.0,40.0)(99.0,40.0)
\Text(95.0,33.0)[]{$\tau_R$}
\Text( 5.0,33.0)[]{$\mu_L$}
\Text(65.0,33.0)[]{$\tilde H^0$}
\Text(35.0,33.0)[]{$\tilde B^0$}
\DashCArc(50.0,40.0)(30.0,0.0,180.0){3.0}
\Text(20.0,70.0)[]{$\tilde \mu_L$}
\Text(83.0,70.0)[]{$\tilde \tau_L$}
\Vertex(50.0,40.0){3}
\Vertex(50.0,70.0){3}
\Text(50.0,82.0)[]{$\delta^{LL}_{23}$}
\end{picture}} \ 
\frac{m_{\tau}}{m_{\mu}} \frac{\partial
{\cal A}^c_{\mu}}{\partial \ln m_{\tilde{\mu}_L}^2} \delta^{LL}_{23} 
\nonumber \\
\parbox{1.5in}{
\begin{picture}(100,60)(0,25)
\Line( 1.0,40.0)(99.0,40.0)
\Text(95.0,33.0)[]{$\tau_R$}
\Text( 5.0,33.0)[]{$\mu_L$}
\Text(65.0,33.0)[]{$\tilde H^0$}
\Text(35.0,33.0)[]{$\tilde W^0$}
\DashCArc(50.0,40.0)(30.0,0.0,180.0){3.0}
\Text(20.0,70.0)[]{$\tilde \mu_L$}
\Text(83.0,70.0)[]{$\tilde \tau_L$}
\Vertex(50.0,40.0){3}
\Vertex(50.0,70.0){3}
\Text(50.0,82.0)[]{$\delta^{LL}_{23}$}
\end{picture}} \ 
&&
\frac{m_{\tau}}{m_{\mu}} \frac{\partial
{\cal A}^d_{\mu}}{\partial \ln m_{\tilde{\mu}_L}^2} \delta^{LL}_{23}
\qquad 
\parbox{1.5in}{
\begin{picture}(100,60)(0,25)
\Line( 1.0,40.0)(99.0,40.0)
\Text(95.0,33.0)[]{$\tau_R$}
\Text( 5.0,33.0)[]{$\mu_L$}
\Text(65.0,33.0)[]{$\tilde H^{\pm}$}
\Text(35.0,33.0)[]{$\tilde W^{\pm}$}
\DashCArc(50.0,40.0)(30.0,0.0,180.0){3.0}
\Text(20.0,70.0)[]{$\tilde \nu_\mu$}
\Text(83.0,70.0)[]{$\tilde \nu_\tau$}
\Vertex(50.0,40.0){3}
\Vertex(50.0,70.0){3}
\Text(50.0,82.0)[]{$\delta^{LL}_{23}$}
\end{picture}} \ 
\frac{m_{\tau}}{m_{\mu}} \frac{\partial
{\cal A}^e_{\mu}}{\partial \ln m_{\tilde{\nu}_{\mu}}^2}\delta^{LL}_{23}
\nonumber \\
\parbox{1.5in}{
\begin{picture}(100,60)(0,25)
\Line( 1.0,40.0)(99.0,40.0)
\Text(95.0,33.0)[]{$\tau_L$}
\Text( 5.0,33.0)[]{$\mu_R$}
\Text(50.0,33.0)[]{$\tilde B^0$}
\DashCArc(50.0,40.0)(30.0,0.0,180.0){3.0}
\Text(20.0,70.0)[]{$\tilde \mu_R$}
\Text(83.0,70.0)[]{$\tilde \tau_L$}
\Vertex(50.0,70.0){3}
\Text(50.0,82.0)[]{$\delta^{RL}_{23}$}
\end{picture}} \ 
&&
{\cal A}^a_{\mu} \frac{1}{\delta^{LR}_{22}} \delta^{RL}_{23} 
\qquad \qquad \ 
\parbox{1.5in}{
\begin{picture}(100,60)(0,25)
\Line( 1.0,40.0)(99.0,40.0)
\Text(95.0,33.0)[]{$\tau_R$}
\Text( 5.0,33.0)[]{$\mu_L$}
\Text(50.0,33.0)[]{$\tilde B^0$}
\DashCArc(50.0,40.0)(30.0,0.0,180.0){3.0}
\Text(20.0,70.0)[]{$\tilde \mu_L$}
\Text(83.0,70.0)[]{$\tilde \tau_R$}
\Vertex(50.0,70.0){3}
\Text(50.0,82.0)[]{$\delta^{LR}_{23}$}
\end{picture}} \ 
{\cal A}^a_{\mu} \frac{1}{\delta^{LR}_{22}} \delta^{LR}_{23} \ ,
\end{eqnarray}
where the amplitudes ${\cal A}^i_{\ell}$ are given in \eqref{dell}.
The branching ratio may then be written as
\begin{equation}
B(\tau\to\mu\gamma)\ =\ {12\pi^3\alpha\over G_F^2 m_{\tau}^2} 
\left[ \left| {\cal M}_L \right|^2 +
       \left| {\cal M}_R \right|^2 \right]
B(\tau\to e\bar{\nu}_e\nu_\tau) \ ,
\label{BR}
\end{equation}
where
\begin{eqnarray}
{\cal M}_L &=& {m_\tau\over m_\mu} 
\left[ {\partial {\cal A}_{\mu}^{\text{tot}} 
\over \partial \ln m_{\tilde\mu_L}^2} 
+ {\partial {\cal A}_{\mu}^{\text{tot}} 
\over \partial \ln m_{\tilde{\nu}_{\mu}}^2}
\right] \delta^{LL}_{23} 
+ {\cal A}^a_{\mu} \frac{1}{\delta^{LR}_{22}} \delta^{LR}_{23} 
\nonumber \\
{\cal M}_R &=& {m_\tau\over m_\mu}
{\partial {\cal A}_{\mu}^{\text{tot}}
\over \partial \ln m_{\tilde\mu_R}^2} 
\delta^{RR}_{23} 
+ {\cal A}^a_{\mu} \frac{1}{\delta^{LR}_{22}} \delta^{RL}_{23} 
\ . \label{epsR}
\end{eqnarray}
Equations (\ref{dell}), (\ref{atot}), (\ref{BR}) and (\ref{epsR})
provide a compact form for the branching ratio for radiative lepton
decays in the mass insertion approximation and are well-suited to
numerical evaluation. (Note that terms subleading in
$m_{\tau}/m_{\mu}$ and linear in the $\delta$s are also easily
computed as ${\cal M}_L = (\partial {\cal
A}_{\mu}^{\text{tot}}/\partial \ln m^2_{\tilde{\mu}_R})
\delta^{RR}_{23}$ and ${\cal M}_R = (\partial {\cal
A}_{\mu}^{\text{tot}}/\partial \ln m^2_{\tilde{\mu}_L})
\delta^{LL}_{23}$, but we neglect them in the analysis to follow.)

\begin{figure}[t]
\begin{minipage}[t]{0.49\textwidth}
\postscript{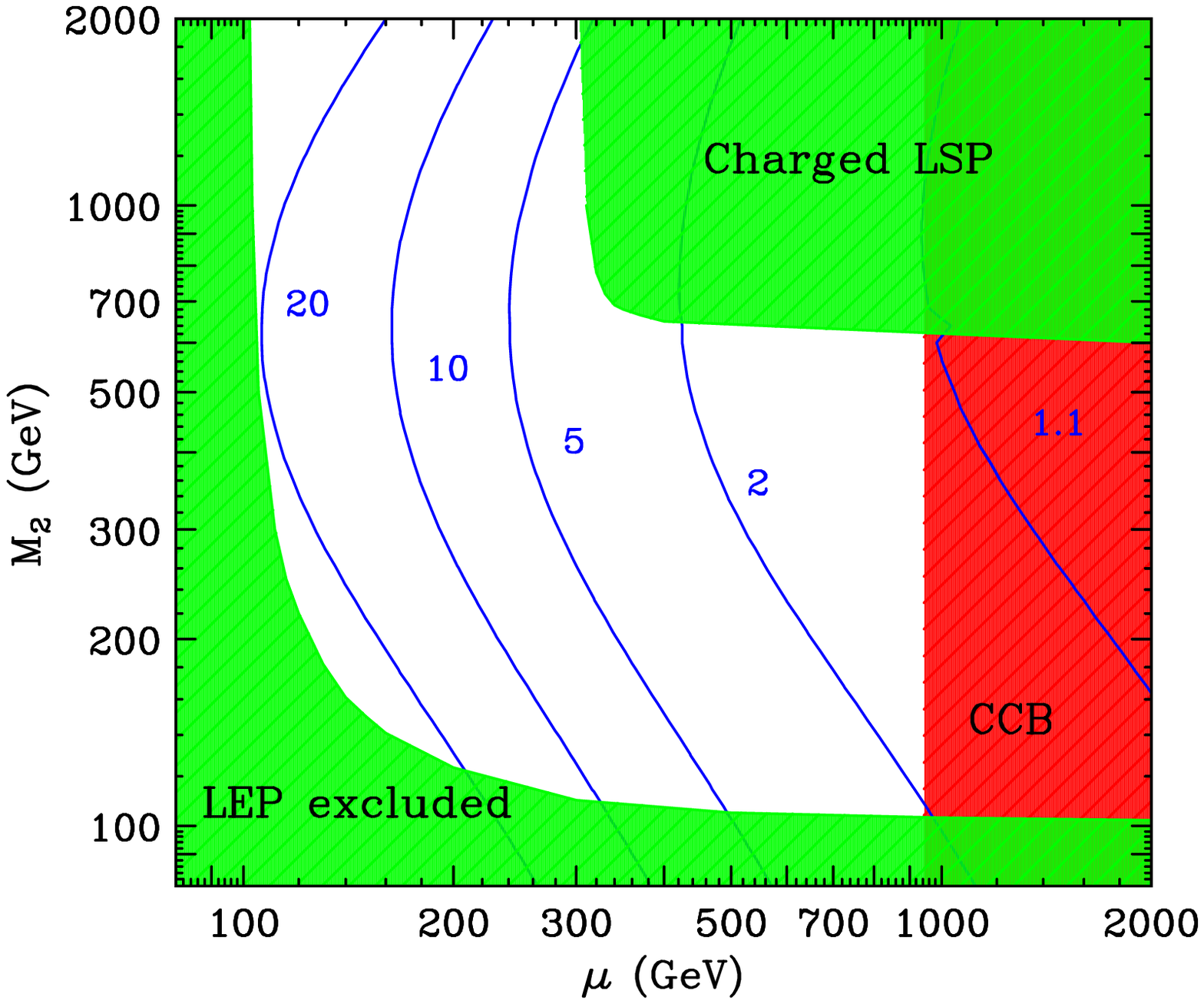}{0.99}
\end{minipage}
\hfill
\begin{minipage}[t]{0.49\textwidth}
\postscript{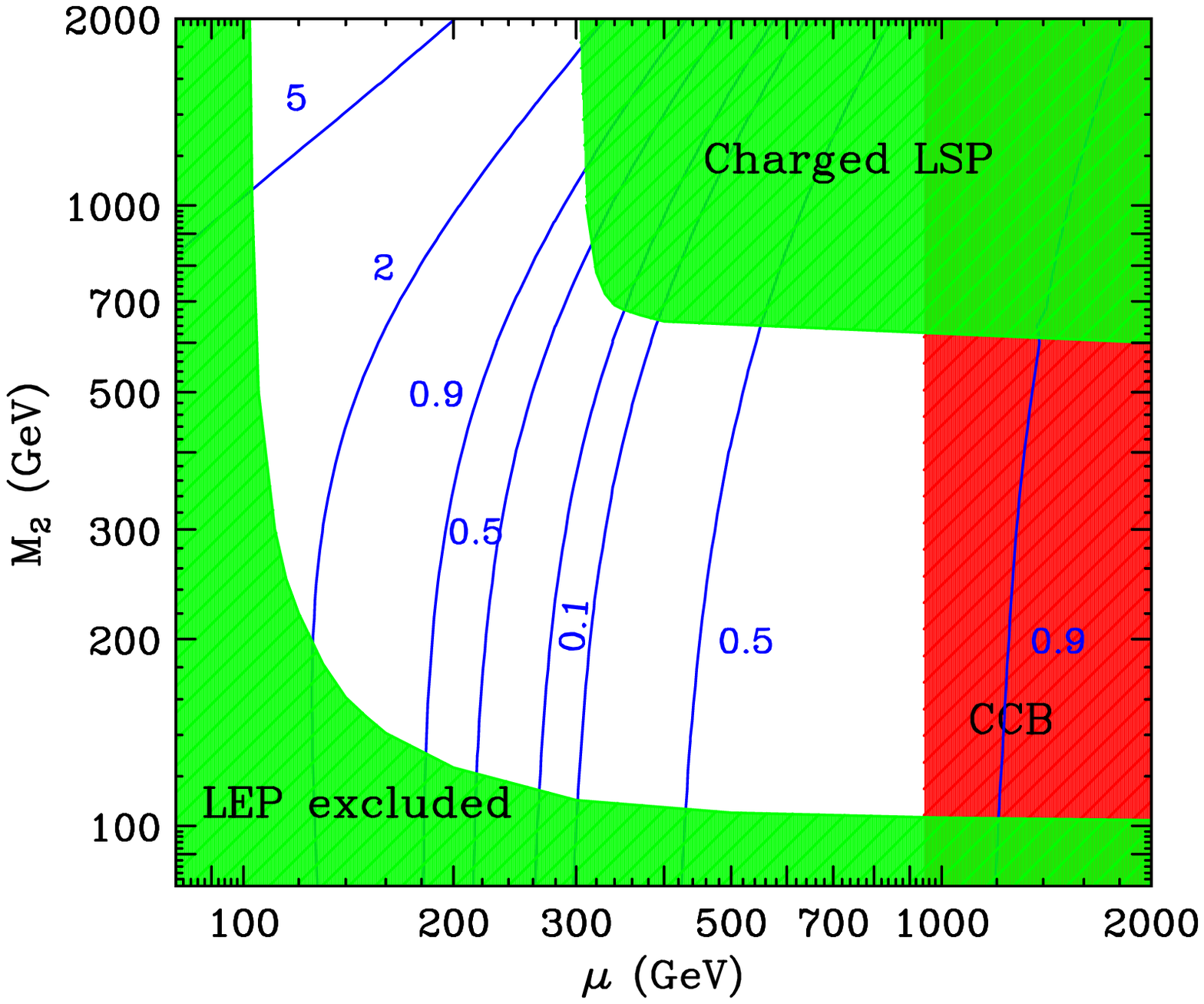}{0.99}
\end{minipage}
\caption{The ratios $\delta^{LL\, \text{max}}_{23\, \Bino\,
\text{only}} / \delta^{LL\, \text{max}}_{23}$ (left) and $\delta^{RR\,
\text{max}}_{23\, \Bino\, \text{only}} / \delta^{RR\,
\text{max}}_{23}$ (right).  The values $\delta^{\text{max}}_{23}$ and
$\delta^{\text{max}}_{23\, \Bino\, \text{only}}$ are the upper bounds
allowed by $\tau \to \mu \gamma$ determined with all leading
contributions and with only the Bino-mediated contribution,
respectively. We fix $m_{\tilde{\ell}} \approx 300~\gev$ and
$\tb=50$. The shaded regions are excluded by LEP, the requirement of a
neutral LSP, and the condition that the vacuum not be charge-breaking.}
\label{fig:comp}
\end{figure}

The flavor-violating mass insertions are bounded by the the current
constraint $B(\tau \to \mu \gamma) < 1.1 \times
10^{-6}$~\cite{Ahmed:2000gh}.  It is important to note that the
Higgsino-mediated decays give the dominant contribution unless $\mu
\gg M_2, M_1$.  The often-used bounds of Gabbiani, Gabrielli, Masiero
and Silvestrini~\cite{Gabbiani:1996hi} assume a photino neutralino,
and so effectively include only the Bino-mediated contribution.  In
Fig.~\ref{fig:comp}, we show contours of the ratio of the upper bound
on $\delta^{LL,RR}_{23}$ determined with only the Bino-mediated
contribution included to the upper bound determined with all of the
leading diagrams included.  We see that the Bino-only bounds are
reasonably accurate only for $|\mu| \agt 1~\tev$, a region that is
forbidden by the requirement that electromagnetic charge be an
unbroken symmetry.  (We have assumed that the diagonal entries of the
stau and smuon mass matrices are equal; in the forbidden region,
$m^2_{\tilde{\tau}_1} <0$.)  Analyses based solely on the Bino
contribution are highly misleading in most regions of parameter space,
especially for moderate and large $\tb$. In particular, for
$\delta^{LL}_{23}$, the constraint from $\tau \to \mu \gamma$ is
always far more stringent than one would conclude from a Bino-only
analysis.  Similar conclusions apply to constraints from $\tau \to e
\gamma$, $\mu \to e \gamma$, and $\mu - e$ conversion and will be
presented elsewhere~\cite{inprep}.

We now determine what effect flavor violation may have on
$\dmu$. Flavor violation induces four $m_{\tau} / m_{\mu}$ enhanced
contributions to the muon's EDM:
\begin{eqnarray}
\parbox{1.5in}{
\begin{picture}(100,60)(0,25)
\Line( 1.0,40.0)(99.0,40.0)
\Text( 5.0,33.0)[]{$\mu_L$}
\Text(95.0,33.0)[]{$\mu_R$}
\Text(50.0,33.0)[]{$\tilde B^0$}
\DashCArc(50.0,40.0)(30.0,0.0,180.0){3.0}
\Text(10.0,50.0)[]{$\tilde \mu_L$}
\Text(90.0,50.0)[]{$\tilde \mu_R$}
\Text(35.0,75.0)[]{$\tilde \tau_L$}
\Text(65.0,75.0)[]{$\tilde \tau_R$}
\Vertex(30.0,62.0){3}
\Vertex(70.0,62.0){3}
\Vertex(50.0,70.0){3}
\Text(15.0,65.0)[]{$\delta^{LL}_{23}$}
\Text(85.0,65.0)[]{$\delta^{RR}_{23}$}
\Text(50.0,82.0)[]{$\delta^{LR}_{33}$}
\end{picture}} \ 
&&
{m_\tau\over m_\mu} 
{\partial^2 {\cal A}^a_{\mu} \over \partial \ln m_{\tilde\mu_L}^2
\partial \ln m_{\tilde\mu_R}^2} \delta^{LL}_{23}\delta^{RR}_{23}
\quad 
\parbox{1.5in}{
\begin{picture}(100,60)(0,25)
\Line( 1.0,40.0)(99.0,40.0)
\Text(95.0,33.0)[]{$\mu_R$}
\Text( 5.0,33.0)[]{$\mu_L$}
\Text(50.0,33.0)[]{$\tilde B^0$}
\DashCArc(50.0,40.0)(30.0,0.0,180.0){3.0}
\Text(15.0,55.0)[]{$\tilde \mu_L$}
\Text(88.0,55.0)[]{$\tilde \mu_R$}
\Text(50.0,80.0)[]{$\tilde \tau_L$}
\Vertex(33.0,64.0){3}
\Vertex(67.0,64.0){3}
\Text(22.0,70.0)[]{$\delta^{LL}_{23}$}
\Text(80.0,70.0)[]{$\delta^{RL}_{23}$}
\end{picture}} \ 
{\delta^{RL}_{23}\over\delta^{LR}_{22}}
{\partial {\cal A}_{\mu}^a \over \partial \ln m_{\tilde\mu_L}^2} 
\delta^{LL}_{23} 
\nonumber \\
\parbox{1.5in}{
\begin{picture}(100,60)(0,25)
\Line( 1.0,40.0)(99.0,40.0)
\Text( 5.0,33.0)[]{$\mu_L$}
\Text(95.0,33.0)[]{$\mu_R$}
\Text(50.0,33.0)[]{$\tilde B^0$}
\DashCArc(50.0,40.0)(30.0,0.0,180.0){3.0}
\Text(10.0,50.0)[]{$\tilde \mu_L$}
\Text(90.0,50.0)[]{$\tilde \mu_R$}
\Text(35.0,75.0)[]{$\tilde \tau_R$}
\Text(65.0,75.0)[]{$\tilde \tau_L$}
\Vertex(30.0,62.0){3}
\Vertex(70.0,62.0){3}
\Vertex(50.0,70.0){3}
\Text(15.0,65.0)[]{$\delta^{LR}_{23}$}
\Text(85.0,65.0)[]{$\delta^{RL}_{23}$}
\Text(50.0,82.0)[]{$\delta^{LR}_{33}$}
\end{picture}} \ 
&&
{m_\tau\over m_\mu} 
{\partial^2 {\cal A}^a_{\mu} \over \partial \ln m_{\tilde\mu_L}^2
\partial \ln m_{\tilde\mu_R}^2} \delta^{LR}_{23}\delta^{RL}_{23}
\quad 
\parbox{1.5in}{
\begin{picture}(100,60)(0,25)
\Line( 1.0,40.0)(99.0,40.0)
\Text(95.0,33.0)[]{$\mu_R$}
\Text( 5.0,33.0)[]{$\mu_L$}
\Text(50.0,33.0)[]{$\tilde B^0$}
\DashCArc(50.0,40.0)(30.0,0.0,180.0){3.0}
\Text(15.0,55.0)[]{$\tilde \mu_L$}
\Text(88.0,55.0)[]{$\tilde \mu_R$}
\Text(50.0,80.0)[]{$\tilde \tau_R$}
\Vertex(33.0,64.0){3}
\Vertex(67.0,64.0){3}
\Text(22.0,70.0)[]{$\delta^{LR}_{23}$}
\Text(80.0,70.0)[]{$\delta^{RR}_{23}$}
\end{picture}} \  
{\delta^{LR}_{23}\over\delta^{LR}_{22}}
{\partial {\cal A}_{\mu}^a \over \partial \ln m_{\tilde\mu_R}^2} 
\delta^{RR}_{23} 
\nonumber \ . 
\end{eqnarray}
The additional flavor-violating contribution to the muon's EDM is then
simply $\dmu^{FV} = \frac{1}{2} e \Im {\cal A}_{\mu}^{FV}$, where
\begin{eqnarray}
{\cal A}_{\mu}^{FV} &=& {m_\tau\over m_\mu} 
{\partial^2 {\cal A}_{\mu}^a \over \partial \ln m_{\tilde\mu_L}^2
\partial \ln m_{\tilde\mu_R}^2} 
\left[ \delta^{LL}_{23}\delta^{RR}_{23} +
       \delta^{LR}_{23}\delta^{RL}_{23} \right] \nonumber \\
&&+ 
{\delta^{RL}_{23}\over\delta^{LR}_{22}}
{\partial {\cal A}_{\mu}^a \over \partial \ln m_{\tilde\mu_L}^2} 
\delta^{LL}_{23} +
{\delta^{LR}_{23}\over\delta^{LR}_{22}}
{\partial {\cal A}_{\mu}^a \over \partial \ln m_{\tilde\mu_R}^2} 
\delta^{RR}_{23} \ .
\label{amuFV}
\end{eqnarray}

\begin{figure}[t]
\begin{minipage}[t]{0.47\textwidth}
\postscript{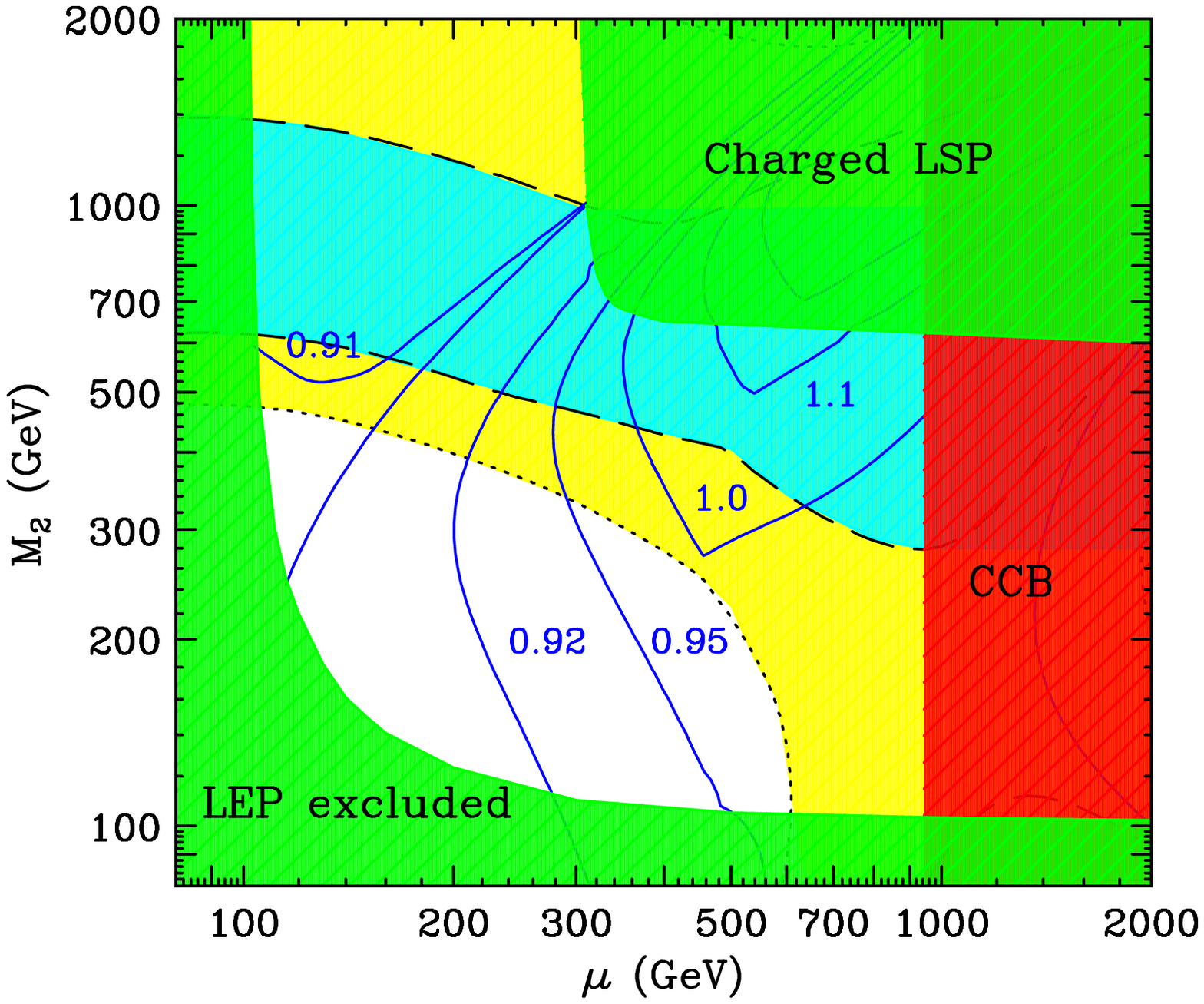}{0.99}
\end{minipage}
\hfill
\begin{minipage}[t]{0.51\textwidth}
\postscript{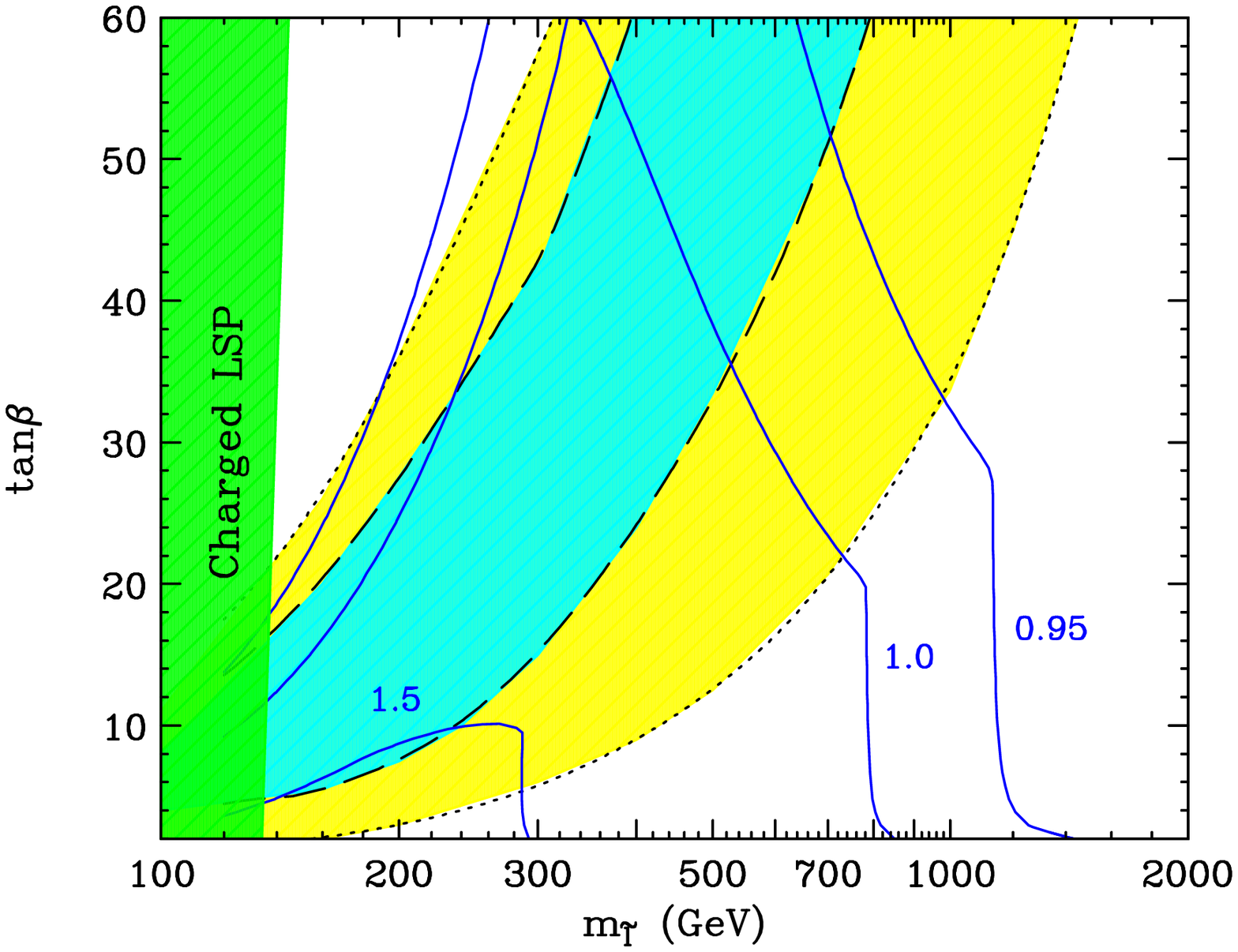}{0.99}
\end{minipage}
\caption{Contours of $\dmu^{\text{max}}$ in units of $10^{-24}~\ecm$
in the presence of flavor violation. Regions consistent with the
observed $\Delta a_\mu$ at the 1$\sigma$ and 2$\sigma$ levels are also
shown. On the left, $m_{\tilde{\ell}} \approx 300~\gev$ and $\tb=50$,
and on the right, $M_2 = 300~\gev$ and $|\mu| = 500~\gev$. $M_1$ is
fixed by gaugino mass unification.  The excluded regions are as in
Fig.~\ref{fig:comp}.}
\label{fig:muM2}
\end{figure}

Contours of the maximal possible EDM $\dmu^{\text{max}}$ in the
presence of $LL$ and $RR$ flavor violation are given in
Fig.~\ref{fig:muM2}.  To obtain $\dmu^{\text{max}}$, $\phi_{\mu}$ is
taken to saturate the constraint from $\de$, and the sign of the
flavor-violating contribution is chosen to add constructively to the
flavor-conserving piece.  The parameters $\delta^{LL,RR}_{23}$ are
also taken to maximize $\dmu$ given the constraint from $\tau \to \mu
\gamma$; we make use of the fact that, subject to the constraint $ax^2
+ by^2 \le c$ with $a,b,c >0$, the product $xy$ is maximized for $ax^2
= by^2 = c/2$.  We also require $\delta^{LL,RR}_{23} \le 1/2$ so that
the mass insertion approximation is valid.  We find that flavor
violation may enhance $\dmu$.  While the enhancement is not enormous,
it does bring the maximal possible value of $\dmu$ into the range of
the proposed experimental sensitivity in parts of the parameter space.

We have performed a similar analysis for chirality-violating flavor
violation.  In the case of non-vanishing $\delta^{LR,RL}_{23}$,
enhancements above the proposed sensitivity are not found.
 
Note, however, that, to investigate various effects independently, we
have assumed real off-diagonal masses.  In fact, however, this
division of new physics effects is rather artificial, as off-diagonal
masses need not be real and generically have ${\cal O}(1)$ phases.

For concreteness, we consider two cases: in the first, we take
$\Arg(\delta^{LL}_{23} \delta^{RR}_{23}) = \phi_{\delta}$ and
$\delta^{LR}_{23} = \delta^{RL}_{23} = 0$, while in the second, we let
$\delta^{LL}_{23} = \delta^{RR}_{23} = 0$ and $\Arg(\delta^{LR}_{23}
\delta^{RL}_{23}) = \phi_{\delta}$.  In both cases, the phase
$\phi_{\delta}$ is irrelevant for $B(\tau \to \mu \gamma)$, as only
the magnitudes of the $\delta$s enter in \eqref{BR}.  This phase is
also not constrained by $\de$, as it has no direct couplings to the
first generation. However, it contributes directly to $\dmu$, as is
clear in \eqref{amuFV}.

\begin{figure}[tbp]
\postscript{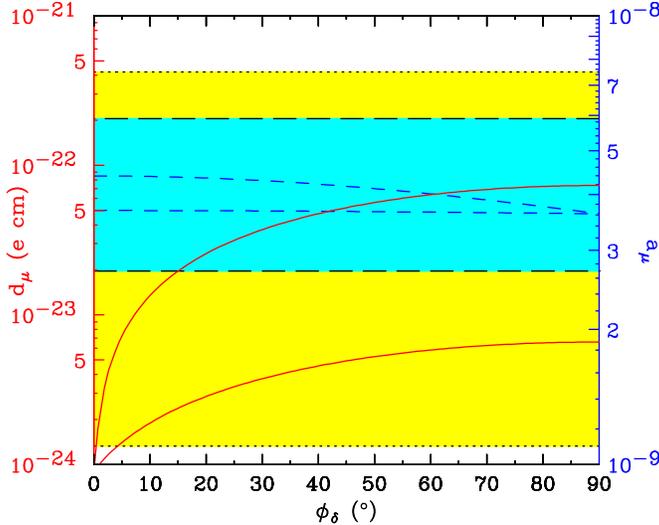}{0.53}
\caption{Contours of $\dmu$ (solid) and $\amu$ (dashed) as functions
of $\phi_{\delta}$, the $CP$-violating phase present in
flavor-violating slepton mass terms. In each pair of contours, the
upper is for $\Arg(\delta^{LL}_{23} \delta^{RR}_{23}) = \phi_{\delta}$
and $\delta^{LR}_{23} = \delta^{RL}_{23} = 0$, while the lower is for
$\delta^{LL}_{23} = \delta^{RR}_{23} = 0$ and $\Arg(\delta^{LR}_{23}
\delta^{RL}_{23}) = \phi_{\delta}$.  We fix $m_{\tilde{\ell}} =
300~\gev$, $\tb = 30$, $|\mu| = 500~\gev$, and $M_2 = 300~\gev$, and
$M_1$ is fixed by gaugino mass unification.}
\label{fig:dmu_thdelta}
\end{figure}

In Fig.~\ref{fig:dmu_thdelta}, we show the dependence of $\dmu$ on the
phase $\phi_{\delta}$ for both cases.  We see that in the $LL/RR$
case, it is easy to achieve values of $\dmu \sim 10^{-22}~\ecm$, two
orders of magnitude above the proposed sensitivity.  For the $LR/RL$
case, $\dmu$ well above the proposed sensitivity is also
possible. Such values are consistent with all present constraints.  In
particular, note that we have also given values of $\amu$ including
both the flavor-conserving contribution and the flavor-violating
amplitude of \eqref{amuFV}.  As may be seen in
Fig.~\ref{fig:dmu_thdelta}, the values of $\phi_{\delta}$ are also
perfectly consistent with the currently preferred $\amu$.

\section{Conclusions}
\label{sec:conclusions}

The proposal to measure the muon EDM at the level of $10^{-24}~\ecm$
potentially improves existing sensitivities by five orders of
magnitude.  Such a leap in sensitivity is rare in studies of basic
properties of fundamental particles and merits attention.

In this study we have considered the muon EDM from a number of
theoretical perspectives.  We noted that the recent results from the
Muon $(g-2)$ Experiment, although widely interpreted as evidence for a
non-standard model contribution to $\amu$, may alternatively be
ascribed entirely to a non-standard model contribution to $\dmu$.  In
fact, these results provide the most stringent constraints on the muon
EDM at present.  Theoretical prejudices aside, this ambiguity will be
definitively resolved only by improved bounds on (or measurements of)
$\dmu$, such as will be possible in the proposed $\dmu$ experiment.

Considering only the indications from a non-standard model
contribution to $\amu$, `naturalness' implies muon EDMs far above the
proposed sensitivity.  In more concrete scenarios, however, additional
constraints, notably from the electron's EDM and lepton flavor
violation, impose important restrictions.  Nevertheless, we have noted
a number of well-motivated possibilities in supersymmetry:
non-degeneracy, non-proportionality, and slepton flavor
violation. Each of these may produce a muon EDM above the proposed
sensitivity.

For simplicity, we have focused for the most part on one effect at a
time.  {}From a model-building point of view, however, this is rather
unnatural.  For example, non-degeneracy of the diagonal elements of
the scalar mass matrices is typically accompanied by flavor violation.
At the very least, if the soft masses are diagonal but non-degenerate
in a particular interaction basis, off-diagonal elements will be
generated upon rotating to the fermion mass basis.  As noted above,
non-proportionality will also typically be accompanied by flavor
violation.  In both of these cases, then, the lepton flavor violation
leads to new contributions to $\dmu$, as well as to new constraints
from lepton flavor violating observables.  The amount of flavor
violation present is highly model-dependent, and so we have not
considered this in detail.  In Sec.~\ref{sec:flavor}, however, we have
noted that $\dmu$ may be greatly enhanced by two or more simultaneous
effects, leading to values of $\dmu \sim 10^{-22}~\ecm$, far above the
proposed sensitivity.

If a non-vanishing $\dmu$ is discovered, it will be unambiguous
evidence for physics beyond the standard model.  At the currently
envisioned sensitivity, it will also imply naive scaling violation,
with important implications for many new physics models.  In addition,
the measurement of non-standard model contributions to both $\dmu$ and
$\amu$ will provide a measurement of new $CP$-violating phases, with
little dependence on the overall scale of the new physics. Such
information is difficult to obtain otherwise. Low energy precision
experiments may not only uncover evidence for new physics before high
energy collider experiments, but may also provide information about
the new physics that will be highly complementary to the information
ultimately provided by colliders.

\section*{Acknowledgments}

We thank R.~Carey, G.~Kribs, J.~Miller, W.~M.~Morse, Y.~Nir, and
B.~L.~Roberts for useful discussions.  This work was supported in part
by the U.~S.~Department of Energy under cooperative research agreement
DF--FC02--94ER40818. KTM thanks the Fermilab Theory Group (summer
visitor program), the Argonne National Lab Theory Group (Theory
Institute 2001), the Aspen Center for Physics, and D.~Rainwater and
M.~Schmaltz for hospitality during the completion of this work.

\section*{Appendix}

The functions $K_N$ and $K_C$ of \eqref{dell} are 
\begin{eqnarray}
K_N(x^2,y^2,z^2) &=&
J_5(x^2,x^2,y^2,z^2,z^2)+J_5(x^2,x^2,y^2,y^2,z^2)\ , \\
K_C(x^2,y^2,z^2) &=& 2I_4(x^2,y^2,z^2,z^2)+2I_4(x^2,y^2,y^2,z^2)
- K_N(x^2,y^2,z^2) \ ,
\end{eqnarray}
where the loop functions $J_n$ and $I_n$ are defined iteratively
\cite{Moroi:1996yh} through
\begin{eqnarray}
I_n(x_1^2,\ldots,x_n^2) &=&
{1\over x_1^2-x_n^2}\ \left[ I_{n-1}(x_1^2,\ldots,x_{n-1}^2)
-I_{n-1}(x_2^2,\ldots,x_n^2)  \right]\ ,\\
J_n(x_1^2,\ldots,x_n^2) &=&
 I_{n-1}(x_1^2,\ldots,x_{n-1}^2)
\ +\ x_n^2\ I_{n}(x_1^2,\ldots,x_n^2) \ ,
\end{eqnarray}
with
\begin{equation}
I_2(x_1^2,x_2^2)\ = \ 
- {1\over 16\pi^2}
\left\{ {x_1^2\over x_1^2-x_2^2}\ln{x_1^2\over\Lambda^2}
+ {x_2^2\over x_2^2-x_1^2}\ln{x_2^2\over\Lambda^2} \right\}\ .
\end{equation}

Their arguments are the gaugino masses $M_1$ and $M_2$, the Higgsino
mass parameter $\mu$, and the scalar soft supersymmetry breaking
masses
\begin{eqnarray}
m_{\tilde{\ell}_L}^2 &=&
({\bold{m}_{LL}^2})_{\ell\ell}+ \left(-{1\over 2} +
\sin^2\theta_W\right) m_Z^2\cos2\beta \nonumber \\
m_{\tilde{\ell}_R}^2 &=&
({\bold{m}_{RR}^2})_{\ell\ell}-\sin^2\theta_W m_Z^2\cos2\beta \\
m_{\tilde{\nu}_{\ell}}^2 &=&
({\bold{m}_{LL}^2})_{\ell\ell}+ \frac{1}{2} m_Z^2\cos2\beta \ . 
\nonumber
\end{eqnarray}



\end{document}